\documentclass[aip,reprint]{revtex4-1}

\usepackage{graphicx}
\usepackage{amssymb}
\usepackage{gensymb}
\usepackage{amsmath} 

\usepackage{xcolor}

\makeatletter
\renewcommand\@make@capt@title[2]{%
 \@ifx@empty\float@link{\@firstofone}{\expandafter\href\expandafter{\float@link}}%
  {\textbf{#1}}\@caption@fignum@sep#2\quad
}%
\makeatother

\draft 

\begin{document}

\preprint{AIP/123-QED}

\title{Atomic-scale mapping and quantification of local Ruddlesden-Popper phase variations}





\author{Erin E. Fleck}
  
\author{Berit H. Goodge}%

\affiliation{ 
School of Applied and Engineering Physics, Cornell University, Ithaca, New York 14853, USA
}%

\author{Matthew R. Barone}
\affiliation{Department of Materials Science and Engineering, Cornell University, Ithaca, NY 14853, USA}

\author{Hari P. Nair}
\affiliation{Department of Materials Science and Engineering, Cornell University, Ithaca, NY 14853, USA}

\author{Nathaniel J. Schreiber}
\affiliation{Department of Materials Science and Engineering, Cornell University, Ithaca, NY 14853, USA}

\author{Natalie M. Dawley}
\affiliation{Department of Materials Science and Engineering, Cornell University, Ithaca, NY 14853, USA}

\author{Darrell G. Schlom}
\affiliation{Department of Materials Science and Engineering, Cornell University, Ithaca, NY 14853, USA}
\affiliation{Kavli Institute at Cornell for Nanoscale Science, Cornell University, Ithaca, NY 14853, USA}
\affiliation{Leibniz-Institut f\"ur Kristallz\"uchtung, Max-Born-Str. 2, 12489 Berlin, Germany}

\author{Lena F. Kourkoutis}
\email{lena.f.kourkoutis@cornell.edu.}
\affiliation{ 
School of Applied and Engineering Physics, Cornell University, Ithaca, New York 14853, USA
}%
\affiliation{Kavli Institute at Cornell for Nanoscale Science, Cornell University, Ithaca, NY 14853, USA}

\date{\today}

\begin{abstract}
The Ruddlesden-Popper ($A_{n+1}B_{n}\text{O}_{3n+1}$) compounds are a highly tunable class of materials whose functional properties can be dramatically impacted by their structural phase $n$.
The negligible energetic differences associated with forming a sample with a single value of $n$ versus a mixture of $n$
makes the growth of these materials difficult to control and can lead to local atomic-scale structural variation arising from small stoichiometric deviations. 
In this work, we present a Python analysis platform to detect, measure, and quantify the presence of different $n$-phases based on atomic-resolution scanning transmission electron microscopy (STEM) images in a statistically rigorous manner. 
We employ phase analysis on the 002 Bragg peak to identify horizontal Ruddlesden-Popper faults which appear as regions of high positive compressive strain within the lattice image, allowing us to quantify the local structure. 
Our semi-automated technique offers statistical advantages by considering effects of finite projection thickness, limited fields of view, and precise sampling rates. 
This method retains the real-space distribution of layer variations allowing for a spatial mapping of local $n$-phases, enabling both quantification of intergrowth occurrence as well as qualitative description of their distribution, opening the door to new insights and levels of control over a range of layered materials.
\end{abstract}


\maketitle

\section{\label{sec:level1}Introduction}
\begin{figure*}
\includegraphics[width=\textwidth]{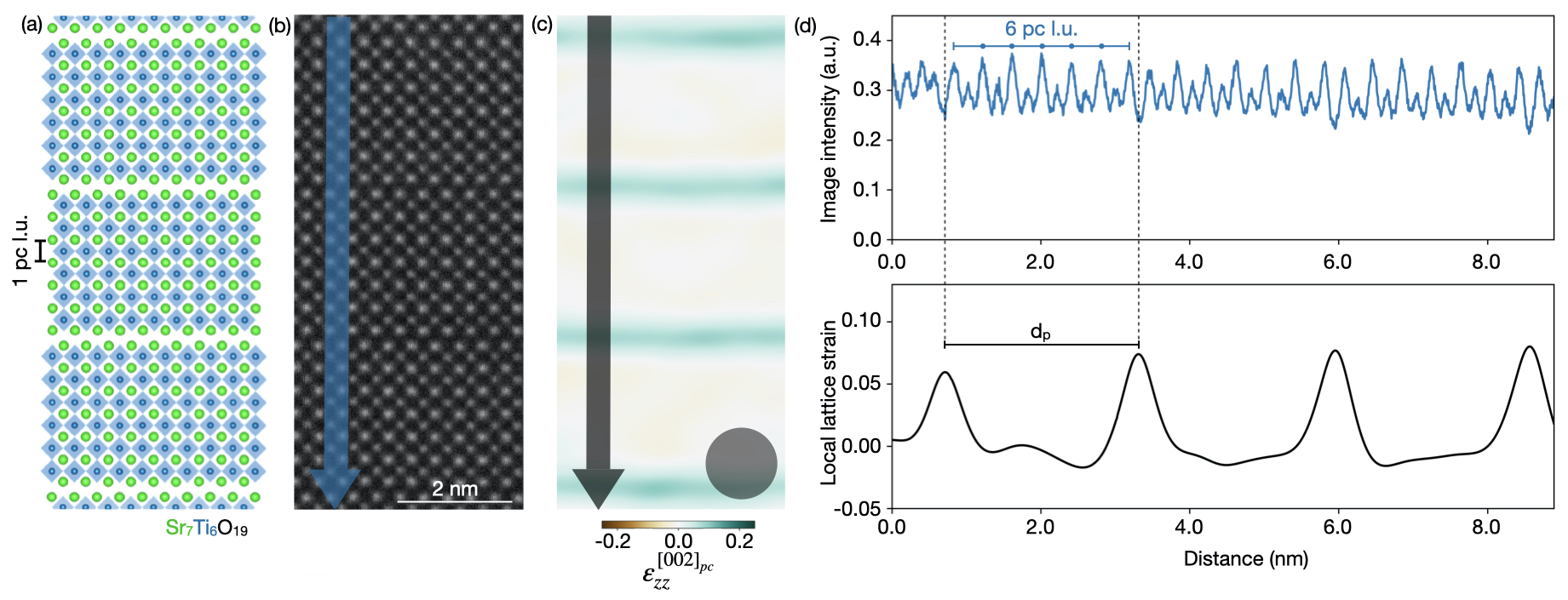} 
\caption{An overview of the counting process for local Ruddlesden-Popper phases. (a) Atomic structure of Sr$_7$Ti$_6$O$_{19}$, an $n$ = 6 Ruddlesden-Popper phase. The pseudocubic lattice unit (pc l.u.) is defined as shown. (b) HAADF-STEM image of Sr$_7$Ti$_6$O$_{19}$ film. (c) Strain map of the same region as (b) generated by lock-in analysis of the 002 Bragg peak. The circle denotes the real-space coarsening length set by the phase lock-in analysis. (d) Top, image intensity averaged over the width of the line shown in (b). Bottom, local lattice strain averaged over the corresponding line in (c). The peak distance, denoted here by d$_p$, corresponds to the distance between two horizontal Ruddlesden-Popper faults.
}
\label{fig*:fig1}
\end{figure*}


Layered materials are characterized by strongly bonded two-dimensional (2D) or quasi-2D atomic planes separated by weaker inter-planar bonding which gives rise to anisoptropic crystal structures and material properties. 
In many cases, the responses of these materials can be tuned by manipulating the dimensionality along the out-of-plane direction.
The Ruddlesden-Popper layered perovskite compounds are one such class, described by the general formula $A_{n+1}B_{n}\text{O}_{3n+1}$, where $A$ is an alkali, alkaline-earth, rare-earth metal, In, Sn, Pb, or Bi and $B$ is a transition metal. \cite{balz1955struktur,ruddlesden1957new,ruddlesden1958compound}
Structurally, $n$ is the number of perovskite $AB$O$_3$ layers separated by a single rock salt $A$O layer, more clearly illustrated by an alternate representation of this formula, ($AB$O$_3$)$_nA$O.
The relative dimensionality of the system is also defined by $n$ such that $n$ = 1 phases are strongly anisotropic while $n$ = $\infty$ are three-dimensional perovskites. 
The atomic structure of an $n$ = 6 Ruddlesden-Popper phase is shown in Fig. \ref{fig*:fig1}a. 
With careful design and growth, variations between these phases can be exploited to yield a wide variety of functional properties, including colossal magnetoresistance, \cite{moritomo1996giant} superconductivity, \cite{muller1989crystal, Pan2021} and dielectric tunability. \cite{lee2013exploiting,dawley2020targeted} 


The growth of Ruddlesden-Popper compounds -- particularly those with higher-$n$ phases -- becomes difficult to control due to the similar stoichiometry and formation energies among nearby members in a given homologous series\cite{barone2021improved} and can lead to the presence of mixed $n$-phase materials.\cite{intergrowth, solarcell, leadhalide} 
For some applications these inclusions may be inert, but in certain cases they may significantly alter the measured response of the material, even for small variations in $n$. 
The Ruddlesden-Popper strontium ruthenates (Sr$_{n+1}$Ru$_n$O$_{3n+1}$), for example, can exhibit remarkably different behaviors at low temperature for small variations in $n$: the $n$ = 1 phase (Sr$_2$RuO$_4$) is an unconventional superconductor, \cite{maeno1994superconductivity, maeno1998enhancement} while higher-$n$ phases ($n = 3$ Sr$_4$Ru$_3$O$_{10}$, $n = 4$ Sr$_5$Ru$_4$O$_{13}$, and $n = \infty$ SrRuO$_3$) are ferromagnetic metals, \cite{crawford2002structure} and the ground state of the intermediate $n$ = 2 phase (Sr$_3$Ru$_2$O$_7$) can be pushed to a ferromagnetic state with very small perturbations. \cite{brodsky2017strain, marshall2018electron} 
Particularly for novel heterostructures that may include one or more such phases carefully chosen for their functional properties, quantifying the occurrence and spatial distribution of different Ruddlesden-Popper layers is therefore of key importance. 

The overall crystallinity of these materials can be assessed with bulk techniques such as x-ray diffraction (XRD), but more precise methods of characterization are required to quantify subtle variations in the precise layering structure. 
Using a sub-\AA \, electron probe, aberration-corrected scanning transmission electron microscopy (STEM) provides a direct, real-space visualization of the atomic lattice and defects or variations therein. 
In high-angle annular dark-field (HAADF) STEM, the relative intensity of each atomic column scales approximately as the square of its atomic number, such that heavier species appear brighter and lighter elements appear dimmer. 
Figure \ref{fig*:fig1}b shows a HAADF-STEM image of the $n$ = 6 Ruddlesden-Popper Sr$_7$Ti$_6$O$_{19}$ in which the layered perovskite motif comprising brighter strontium ($Z = 38$) and dimmer titanium ($Z$ = 22) atomic columns is clearly visible, as are adjacent rock salt spacer layers at the horizontal Ruddlesden-Popper faults of each perovskite slab.
By imaging these layered materials in cross-section, STEM provides a platform to directly identify the occurrence of $n$-phase variations with exceptional sensitivity. 
In order to build an accurate representation of a macroscopic sample, however, it is imperative to establish a systematic and statistical method for quantifying such variations.

Here, we present an openly available analysis platform implemented in Python that can detect, measure, and quantify the presence of different Ruddlesden-Popper $n$-phases based on atomic-resolution STEM images in a statistically rigorous manner. 
We employ phase lock-in analysis \cite{Goodge_phase} to identify the expanded interplanar distance at horizontal Ruddlesden-Popper faults, which appear as regions of high tensile strain within the lattice image. 
We use these high-strain faults to identify and quantify the local $n$-phases within a given image. 
Repeating the analysis over multiple datasets thus provides a statistical characterization of a sample, providing both a quantitative measure of $n$-phase occurrences as well as real-space mapping of their distribution within the sample. 


\section{Methods}
\subsection{Data collection}
Ruddlesden-Popper thin films were grown by molecular beam epitaxy (MBE) as described previously. \cite{nair2018demystifying, dawley2020defect, barone2021improved}
Cross-sectional STEM specimens were prepared using the standard focused ion beam (FIB) lift-out process on a Thermo Scientific Helios G4 UX FIB or FEI Strata 400 FIB equipped with an Omniprobe AutoProbe 200 nanomanipulator. HAADF-STEM images were acquired on an aberration-corrected FEI Titan Themis operating at 120 kV (Sr$_7$Ti$_6$O$_{19}$) or 300 kV (Sr$_2$RuO$_{4}$ and (Sr$_{0.4}$Ba$_{0.6}$)$_{21}$Ti$_{20}$O$_{61}$) with probe convergence semi-angles of 21 and 30 mrad, respectively.

\begin{figure*}
\includegraphics[width=\textwidth]{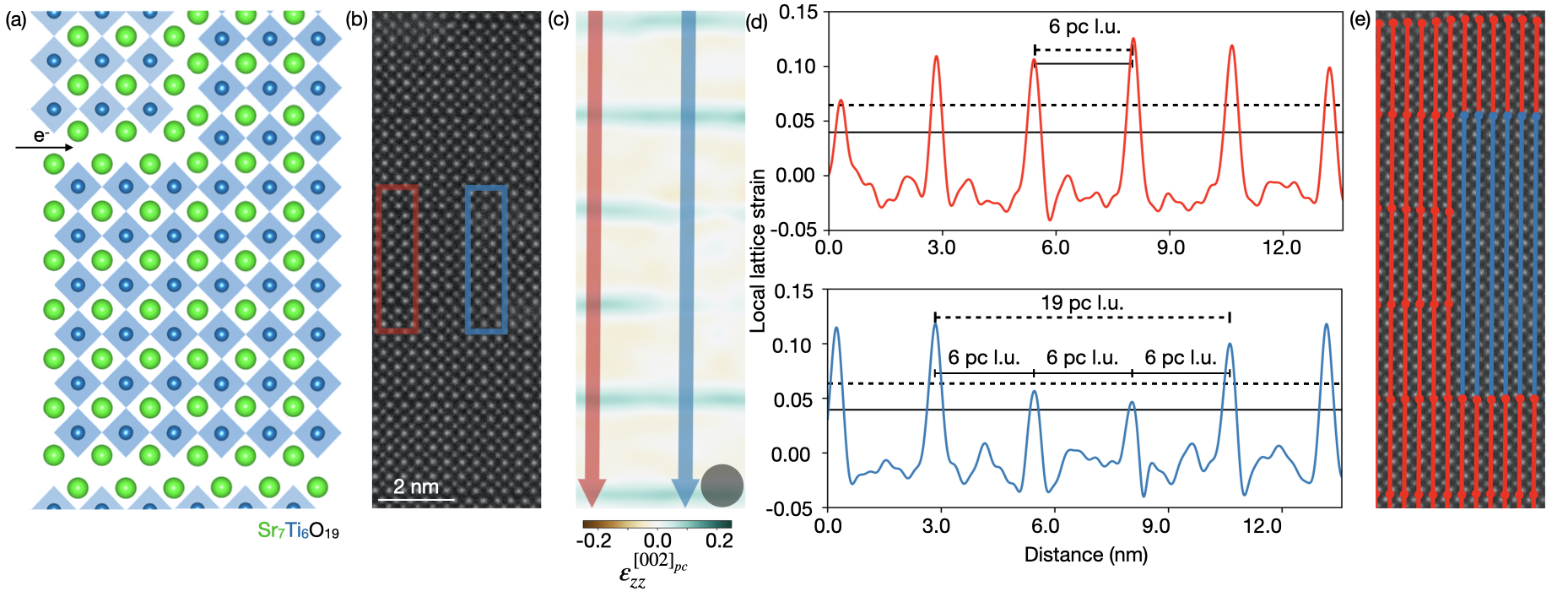}
\caption{Consequences of STEM imaging projection through cross-sectional specimens of finite thickness. (a) A visualization of the electron beam encountering a stacking fault in a Ruddlesden-Popper phase during STEM imaging, which could give rise to blurred contrast as shown in (b), a HAADF-STEM image of an Sr$_7$Ti$_6$O$_{19}$ film. The red box highlights a region of the film with distinct rock salt spacer layers, while the blue box highlights the region on the right where contrast in the horizontal Ruddlesden-Popper fault has blurred due to mixed projection. (c) Tensile strain map of the same region as (b) generated by lock-in analysis of the 002$_{pc}$ Bragg peak, with the circle again denoting the real-space coarsening length set by phase lock-in analysis. (d) Line profiles of local lattice strain averaged over widths of the red (above) and blue (below) arrows in (c). The solid and dashed lines represent different strain thresholds and the pseudocubic  lattice unit counts (pc l.u.) resulting from each. (e) Map of local Ruddlesden-Popper phases in the region shown in (b), where regions of $n$ = 6 and $n$ = 19 are indicated by red and blue lines, respectively. }
\label{fig*:fig2}
\end{figure*}

\subsection{Fourier analysis}
Fourier analysis is an effective way to extract detailed structural information from the periodic frequencies of STEM images.
In images of crystalline materials, the periodicities of the lattice structure give rise to strong peaks in the FFT at spatial frequencies which correspond to the distances between high-symmetry planes. 
Isolating the frequency associated with a single peak provides a tool to extract the image contributions from a chosen set of atomic lattice planes (fringes). 
Further analysis can extract subtle changes to the periodicities in a lattice image, such as variations in the spacing between atomic planes.\cite{hytch1998, Goodge_phase}

Here, we identify the rock salt spacer layers between adjacent perovskite slabs by extracting maps of local lattice strain measured from the 002$_{pc}$ (pseudocubic) peak, which is sensitive to both the $A$O and $B$O$_2$ planes along the growth direction of the film. 
Strain analysis on these lattice fringes highlights the horizontal Ruddlesden-Popper faults due to the difference in interplanar spacing between consecutive $A$O-$A$O layers at the horziontal faults and $A$O-$B$O$_2$ layers within a perovskite slab. 
For example, in perovskite-phase SrTiO$_3$ the pseudocubic $c$-axis lattice constant is 3.905 Å, so the (002) plane spacing is $\sim$3.905 Å / 2 $\approx$ 1.95 Å. 
At the rock salt spacer layers between adjacent $n$ = 6 strontium titanate Ruddlesden-Popper phases, however, the spacing between consecutive SrO planes is 2.79 Å. 
Thus, a frequency-based analysis on this peak will highlight horizontal Ruddlesden-Popper faults as strong interplanar expansion from $\sim$1.95 Å to 2.79 Å.
This expansion appears as local tensile lattice strain, which yields out-of-plane strain maps like the one shown in Fig. \ref{fig*:fig1}c. 
We note that while similar analysis performed on the 001 peak similarly identifies in-plane Ruddlesden-Popper faults, it is less suited to the methods described here because it also highlights in-plane anti-phase horizontal faults where $B$O$_2$ planes are offset along the $c$-axis (vertical Ruddlesden-Popper faults) and regions of blurred contrast arising from mixed atomic projection in the cross-sectional specimen. 
The general technique presented here can be further extended to other Fourier peaks depending on the analytical applications of interest such as isolating only vertical faults with the 200$_{pc}$ peak or cleanly identifying both horizontal and vertical faults using the 101$_{pc}$ peak.


It is additionally important to note that what we refer to as ``strain maps'' are measuring local changes to the interplanar spacing or periodic fringes, which include both real crystalline lattice strains as well as unstrained effects such as the expansion across the horizontal Ruddlesden-Popper fault. 
Additionally, because this technique is a relative measurement the strain across a single image must average to zero such that the small apparent negative strain in the bulk-like regions compensates for the very strong positive strain that appears at the horizontal Ruddlesden-Popper faults. 
The pure-perovskite regions of the sample thus appear to show small negative strain signatures although there is no change in interplanar spacing within these regions.

\subsection{Measurement}
The local 002$_{pc}$ tensile strain map shown in Fig. \ref{fig*:fig1}c demonstrates how this analysis generates signatures of high positive strain directly at horizontal Ruddlesden-Popper faults. 
We simplify the task of identifying these faults by converting the two-dimensional image of local lattice strain into a series of line profiles of strain intensity along the crystalline $c$-axis.
Local maxima in the strain profile identify the vertical position of each fault. 
As demonstrated in Fig. \ref{fig*:fig1}d, the real-space distance between two consecutive peaks will directly correspond to the distance between two rock salt spacer layers, denoted here as $d_p$.
The measurement of this distance in image pixels can be converted into the local $n$-phase using a calibration of the lattice constant determined by the FFT. 

Because the peak within each line profile falls in the middle of a horizontal Ruddlesden-Popper fault rather than centered on either of the $A$O planes, the precise measurement between two strain peaks is a non-integer value of pseudocubic (pc) lattice units (l.u.) which overestimates the local $n$-phase.
This extra distance can be seen in Fig. \ref{fig*:fig1}d, where the peaks in the local lattice strain profile fall slightly outside the STEM image intensity profile peaks which correspond to the first and last atomic planes of a Ruddlesden-Popper layer. 
This excess, however, will necessarily not exceed one pseudocubic lattice unit because the spacing between rock salt spacer layers is smaller than the pseudocubic lattice constant. 
The integer corresponding to the local $n$-phase is then obtained by flooring the calculated number of pseudocubic lattice units.

Repeating this measurement process for numerous line profiles across an image yields quantitative data about both the number of occurrences of different $n$-phases within a sample as well as their spatial distribution. 
In the following sections we discuss some of these effects and how careful applications of this technique can yield new insights to these kinds of layered materials.

\begin{figure*}
\includegraphics[width=\textwidth]{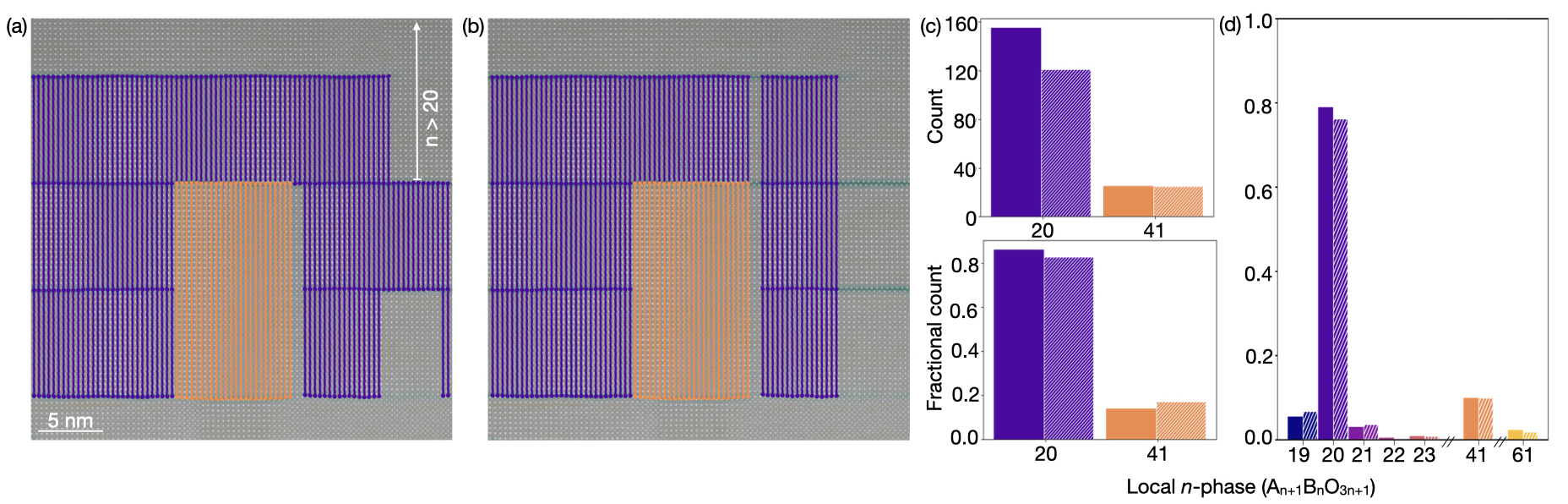}
\caption{An application of statistical corrections to avoid bias towards smaller $n$-phases on an (Sr$_{0.4}$Ba$_{0.6}$)$_{21}$Ti$_{20}$O$_{61}$ film. (a) Mapping of local Ruddlesden-Popper phases as counted before statistical corrections and (b) after statistical corrections are applied. (c) Top, counts of local Ruddlesden-Popper phases without statistical corrections (solid) and after (hashed). Bottom, the same data displayed as a fraction of total occurrences. (d) The fractional count before (solid) and after (hashed) statistical corrections are applied across a series of several STEM images from different regions of the film (reproduced from \cite{barone2021improved}).}
\label{fig*:fig3}
\end{figure*}

\subsection{Projection effects}
The fidelity of this strategy relies on the ability to clearly identify horizontal Ruddlesden-Popper faults as local maxima in crystalline strain profiles. 
Therefore, additional consideration is required to account for the finite projection thickness inherent to cross-sectional STEM imaging.
Figure \ref{fig*:fig2}a shows a schematic side view of how the electron beam may encounter stacking faults as it passes through the sample. 
In projection, these stacking faults create regions where two or more Ruddlesden-Popper structures are vertically offset from each other by half of a perovskite lattice unit in the plane of the specimen due to an ($A$O)$_2$ Ruddlesden-Popper fault. 
This mixed projection through both $A$- and $B$-site atomic columns results in reduced contrast in HAADF-STEM imaging and obscures the clarity of horizontal Ruddlesden-Popper faults compared to crystallographically clean regions of the film. 

Figures \ref{fig*:fig2}b and \ref{fig*:fig2}c show a region of an $n$ = 6 Ruddlesden-Popper Sr$_7$Ti$_6$O$_{19}$ film that contains faults which are clean in some portions (red), but are more blurred elsewhere (blue) such that it is not clear if the region in the blue box contains a complete rock salt spacer layer. 
In the red region, clear atomic contrast between $A$ and $B$ sites (here, Sr and Ti) and distinct horizontal Ruddlesden-Popper faults are visible in the HAADF-STEM image. 
A strain profile across faults in this region contains obvious strong peaks which can be automatically identified as horizontal Ruddlesden-Popper faults by their intensities, widths, and spacing (Fig. \ref{fig*:fig2}d, top). 

Elsewhere, such as in the blue box, the HAADF-STEM image shows faults which do not extend through the full projection thickness of the STEM lamella ($\sim$20-30 nm), similar to the schematic in Fig. \ref{fig*:fig2}a.
The corresponding strain profile across this region, shown in Fig. \ref{fig*:fig2}d (bottom), has peaks of intermediate intensity clearly above the background noise of the strain maps but significantly lower than at clean horizontal faults. 
Including or excluding these peaks in the automated Ruddlesden-Popper analysis will give rise to different interpretations of the local $n$-phase. 
To account for this in our analysis, a user-defined threshold mandates a minimum peak height to qualify as a ``sufficient'' horizontal Ruddlesden-Popper fault for structural characterization. 

For example, in Fig. \ref{fig*:fig2}d (bottom), choosing a local lattice strain threshold of $<$0.05 lends itself to a generous interpretation of a sufficient horizontal Ruddlesden-Popper fault, classifying this region of the film as three distinct local $n$ = 6 phases. 
A higher user-defined threshold (local lattice strain $\geq$0.05) more strictly enforces that sufficient horizontal Ruddlesden-Popper faults extend through the full lamella thickness, classifying regions of mixed projection as the measured $n$-phase between ``complete'' faults. 
Here, the three adjacent $n$ = 6 layers identified by the lower tolerance are instead counted as one singular $n$ = 19 phase (note that the spacing between rock salt spacer layers at a horizontal Ruddlesden-Popper fault allows for one additional ($A$O)$_2$ plane, resulting in $n$ = 19 rather than $n$ = 6 $\times$ 3 = 18). 

Enabling the user to define the threshold for a horizontal Ruddlesden-Popper fault allows for unique standardization of the quantification process for each sample and application. 
Appropriate selection of this threshold is particularly important for samples with significant regions of mixed projection, while for cleaner systems variable threshold values will yield identical quantitative results (Fig. \ref{fig*:fig2}d, top). 
Experimentally, the effects due to the projection nature of STEM imaging can be minimized by careful preparation of high-quality, thin STEM cross-sections.

The same strain profile peaks which are used to extract the $n$-phase can be correlated to real-space image coordinates $xy$ which correspond to the horizontal Ruddlesden-Popper faults identified for each line of analysis. 
We use these coordinates to ``map'' local $n$-phases represented by differently colored lines, e.g., red for $n$ = 6 and blue for $n$ = 19 in Fig. \ref{fig*:fig2}e.
This provides a useful visualization tool for statistical analysis as well as detailed qualitative insights of sample growth or heterostructuring.


\subsection{Statistical considerations}


One consideration that further complicates this characterization process is the limited field of view of each image. 
Unlike in the $n$ = $\infty$ end-member of the Ruddlesden-Popper series, the stacking sequence of finite-$n$ Ruddlesden-Popper phases means that imaging and characterization of a sample is not translationally invariant. 
A naive application of a counting process with this limitation can result in data bias when the edges of a particular Ruddlesden-Popper region fall outside the field of view.
This bias increases when the characteristic distance between horizontal Ruddlesden-Popper faults is larger, i.e., for samples with greater prevalence of high-$n$ phases. 
Figure 3 demonstrates how the limited field of view can impact statistical results in an $n$ = 20 Ruddlesden-Popper (Sr$_{0.4}$Ba$_{0.6}$)$_{21}$Ti$_{20}$O$_{61}$ film. 
The upper-right corner of the field of view shown in Figs. 3a and 3b contains a region with local $n$ $>$ 20. 
Characterization of other images of the same sample suggests that this region is likely $n$ = 41 or $n$ = 61. 
From a single image with this limited field of view, however, determining the exact $n$-phase along line segments like the one labeled ``$n \, >$ 20'' in Fig. 3a is not possible because larger $n$ phases are more likely to extend outside the image field of view. 
We additionally discard all other segments along the same vertical line profile to prevent a statistical bias towards smaller $n$-phases.
By requiring each full line profile used for Ruddlesden-Popper analysis to ``start'' and ``end'' on or near the same rock salt spacer layer, we limit our quantitative analysis to only areas which can be fully described without any regions of uncertain $n$.

\begin{figure*}
\includegraphics[width=\textwidth]{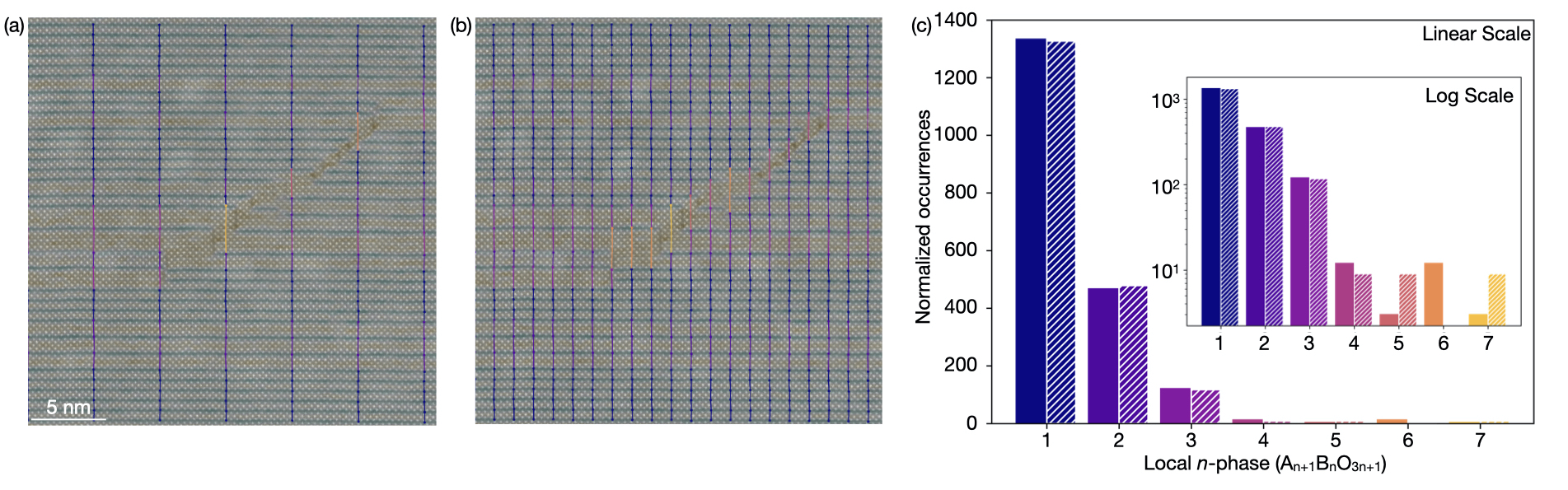} \caption{HAADF-STEM image of nominally $n$ = 1 Ruddlesden-Popper Sr$_2$RuO$_4$ with overlays of the local lattice strain map and lines color-coded to represent the identified local $n$-phase with a horizontal sampling rate of every (a) 10 atomic columns and (b) 3 atomic columns. (c) Statistical counts of local Ruddlesden-Popper phase occurrences from (a, solid) and (b, hashed) normalized to an equivalent sampling of every atomic column.}%
\label{fig*:fig4}
\end{figure*}

As illustrated in Fig. \ref{fig*:fig3}, applying this correction to a single image of an $n$ = 20 Ruddlesden-Popper film shows only a slight change in the relative counts of local $n$-phases. 
The total counts of the $n$ = 20 phase (Fig. \ref{fig*:fig3}c, top) decrease from the uncorrected to the corrected analysis in which weak and out-of-view horizontal faults have resulted in discarded data, such as in the region on the right of Fig. \ref{fig*:fig3}b.
When renormalized by the total number of counts in each case, the effect on fractional counts is much less severe (Fig. \ref{fig*:fig3}c, bottom). 
Across a larger dataset comprising several images of the same sample, shown in Fig. \ref{fig*:fig3}d, a similar trend is observed in which the corrected analysis shows a small reduction in the fractional counts of $n$ = 20 regions.

An additional consideration is the horizontal sampling rate of the analysis. 
The real-space horizontal sampling distance is calibrated from the FFT in a similar way to the vertical lattice plane distance described above. 
The most accurate analysis would have a sampling rate of every atomic column, as shown in Figs. \ref{fig*:fig3}a and \ref{fig*:fig3}b. 
Sampling at this high frequency, however, is computationally expensive.
Further analysis reveals that sampling at much lower frequencies can be a viable and more efficient method of obtaining statistical characterization of a sample from several data sets covering a larger total field of view. 
The maps of local Ruddlesden-Popper phases shown in Figs. \ref{fig*:fig4}a and \ref{fig*:fig4}b have sampling rates of every 10 and 3 atomic columns, respectively. 
Despite the sparser sampling rate shown in Fig. 4a compared to that in Fig. 4b, when the total counts are normalized to a sampling rate of every atomic column as shown in Fig. \ref{fig*:fig4}c, there is very little difference in the distribution of dominant $n$-phases. 
For lower $n$-phases ($n \leq$ 3), both the linear and the inset log scale in Fig. 4c show similar counts for the sparser (solid) and less sparse (hashed) sampling. 
With a decreased sampling rate there is some decrease in the precision of the $n$-phase quantification, particularly for infrequently occurring $n$-phases. 
In this nominally $n$ = 1 Sr$_2$RuO$_4$ sample, the higher $n$-phases (particularly $n \geq$ 4) are counted at different relative occurrences for the two sampling rates, most clearly visible in the inset log scale graph in Fig. 4c. 
Most generally, we expect that a denser sampling rate will more precisely characterize the structural phases of a given sample. 
Here, however, we note that the sparser sampling rate identifies a local $n$ = 6 phase while the denser sampling does not.
Given the relatively low overlap between the chosen sampling frequencies (every 3 and 10 atomic columns), the two analyses yield subtly different quantitative results while reflecting a similar qualitative picture. 
Sparse sampling down to 10$\times$ less than the ideal therefore offers a relatively fast, computationally inexpensive way to characterize a sample's microstructure with little loss in accuracy for the dominant $n$-phases. 


\begin{figure*}
\includegraphics[width=\textwidth]{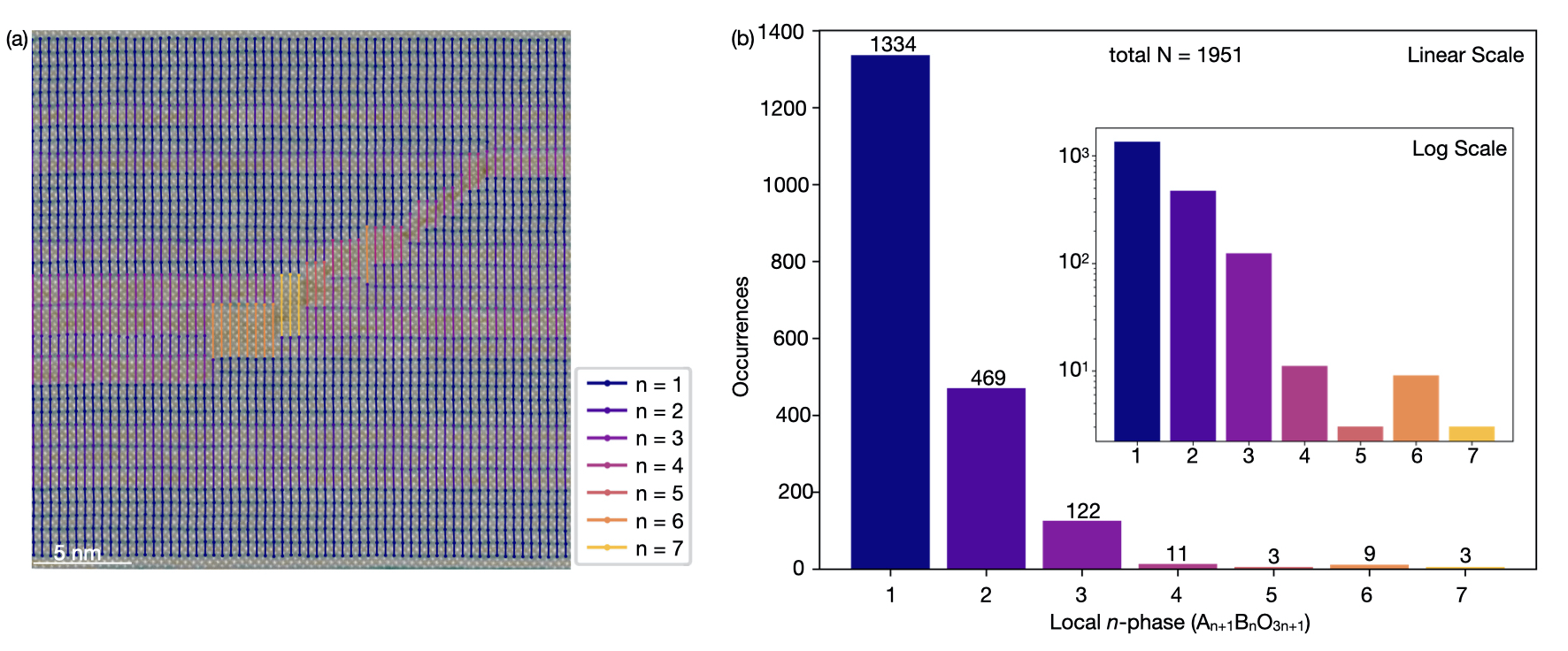} 
\caption{Final results of an ideal sampling mapped onto the same HAADF-STEM image as in Fig. 4. (a) HAADF-STEM image of nominally $n$ = 1 Ruddlesden-Popper Sr$_2$RuO$_4$ with overlays of the local lattice strain map and lines color-coded to represent the identified local $n$-phase. (b) Statistical counts of local Ruddlesden-Popper phase occurrences in (a) plotted on linear and (inset) log scale. }
\label{fig*:fig5}
\end{figure*}

\section{Discussion}
In addition to quantifying different local $n$-phases within a sample, perhaps the most powerful capability of this analysis platform is visually mapping their spatial distribution. 
Figure \ref{fig*:fig5}a contains an ideal analysis (i.e., sampling every atomic column) of the same nominally $n$ = 1 Sr$_2$RuO$_4$ film as in Fig. \ref{fig*:fig4}. 
The histograms of local $n$-phases in Fig. \ref{fig*:fig5}b show qualitative agreement with those in Fig. 4c, as expected. 

Deeper insight to the sample growth process, however, can be gleaned from mapping these phase variations in real space. 
For example, within this film the regions of higher $n$-phases are clustered together towards the center of Fig. \ref{fig*:fig5}a. Beyond this small diagonal cluster, the remainder of the film visible in this field of view is primarily comprised of local regions of $n$ = 1 and 2, with a few inclusions of $n$ = 3. 
In this sample, which was grown by adsorption-controlled MBE,\cite{nair2018demystifying} the local Ruddlesden-Popper variations reflect local variations in elemental stoichiometry likely related to fluctuating chamber conditions described by the windows of thermodynamic stability for each local $n$-phase. \cite{nair2018synthesis, Goodge2022}
For materials grown by shuttered deposition or other sequential approaches, we can similarly extract information about precise synthesis conditions from variations within the sample. 
Real-spacing mapping of local $n$-phase distributions thus provides insight into spatial and temporal fluctuations in a material and its synthesis, which can be utilized to modify the design and growth process to minimize or exploit defects and intergrowths. 
Such analysis will be particularly critical for the design and realization of higher $n$-phase Ruddlesden-Poppers, given that growth is increasingly difficult to control with increasing $n$. 

At a smaller scale, the atomic reconstructions that form between regions of dissimilar $n$ may also have an impact on the local behavior of these layered materials.
This real-space visualization of the distribution of local $n$-phases also highlights faults between regions of different local $n$-phases.
As previously noted in the discussion of Fig. \ref{fig*:fig2}d, due to the difference in interplanar spacing for $A$O-$A$O layers and $A$O-$B$O$_2$ layers, ``missing'' horizontal Ruddlesden-Popper faults do not simply generate a local $n$-phase twice that of the primary phase.
Instead, the resulting fault where dissimilar phases meet may result in a local out-of-plane strain. 
While most thin film growth is concerned with the in-plane epitaxial lattice strain, in certain cases the local out-of-plane strain imposed by heterogeneity in the $n$-phase may play an important role. \cite{moore2021charged}
In other cases, layers of mixed $n$ can be harnessed to help alleviate substrate effects such as step edges.\cite{kim2021superconducting} 
Thus, in regions of high heterogeneity, mapping the junctions between various $n$-phases can provide insight to how local structural phases are formed and arranged within the material.

Understanding the competition or interaction between different phases is a key question for a wide array of Ruddlesden-Popper systems. 
In layered nickelates, for example, where the structural Ruddlesden-Popper phase governs the formal nickel valence \cite{NickelateDiscovery, Pan2022, Pan2021}, this technique is a practical way of mapping where local electronic changes may occur. 
Additionally, certain engineered superlattices comprised of layers with alternating $n$-phases have been shown to host Heisenberg antiferromagnetism.\cite{Kim2022} 
The method presented here allows for both quantification and visualization of the syntactic phases in Ruddlesden-Popper (heterostructures) and other layered compounds, paving the way to a deeper understanding of how such variations interact and interface with each other.  

\begin{acknowledgments}
E.E.F., B.H.G., H.P.N., N.J.S., D.G.S., and L.F.K. acknowledge support by the National Science Foundation Platform for Accelerated Realization, Analysis, and Discovery of Interface Materials (PARADIM) under Cooperative Agreement No. DMR-1539918.
The synthesis science work of M.R.B., N.M.D. and D.G.S. was supported by the U.S. Department
of Energy, Office of Basic Sciences, Division of Materials Sciences and Engineering, under Award
No. DE-SC0002334. This research was funded in part by the Gordon and Betty Moore
Foundation's EPiQS Initiative through Grant No. GBMF9073 to Cornell University.
This work made use of the Cornell Center for Materials Research (CCMR) Shared Facilities, which are supported through the NSF MRSEC Program (No. DMR-1719875). The FEI Titan Themis 300 was acquired through No. NSF-MRI-1429155, with additional support from Cornell University, the Weill Institute, and the Kavli Institute at Cornell. The Thermo Fisher Helios G4 UX FIB was acquired with support by NSF No. DMR-1539918. 
\end{acknowledgments}

\section*{Author Declarations}
\subsection*{Conflict of Interest}
The authors have no conflicts to disclose.
\section*{Data Availability}
The data that support the findings of this study as well as the developed python code have been deposited in the Platform for the Accelerated Realization, Analysis, and Discovery of Interface Materials (PARADIM) database and are available at https://doi.org/XXXX.

\nocite{*}
\bibliography{references}

\end{document}